\documentclass[%
 aip,
 jmp,%
 amsmath,amssymb,
preprint,%
]{revtex4-1}

\usepackage{graphicx}
\usepackage{dcolumn}
\usepackage{bm}

\begin{document}

\title[]{Crumpling for Energy: Modeling Generated Power from the Crumpling of Polymer Piezoelectric Foils for Wearable Electronics}
\thanks{Corresponding Author: Sanjiv Sambandan, Asst. Professor, Dept. of Instrumentation and Appl. Phys., Indian Institute of Science and Lecturer, Solid State Electronics and Nanoscale Science Group, Dept. of Engg., University of Cambridge; email: ssanjiv@isu.iisc.ernet.in, ss698@cam.ac.uk. Sanjiv Sambandan thanks the University of Cambridge, the Indian Institute of Science and the Department of Biotechnology, India for permitting a joint appointment via the DBT-Cambridge Lectureship Program.}

\author{Prakash Kodali}
\affiliation{Department of Instrumentation and Applied Physics, Indian Institute of Science, Bangalore-560012, India}

\author{Ganapathy Saravanavel}%
\affiliation{Department of Instrumentation and Applied Physics, Indian Institute of Science, Bangalore-560012, India}

\author{Sanjiv Sambandan}
\affiliation{Department of Instrumentation and Applied Physics, Indian Institute of Science, Bangalore-560012, India}
\affiliation{Solid State Electronics and Nanoscale Science Group, Department of Engineering, University of Cambridge, Cambridge, UK, CB3 0FF}

\begin{abstract}
We consider possibility of embedding large sheets of polymer piezoelectrics in clothing for sensing and energy harvesting for wearable electronic applications. Power is generated by the crumpling of clothes due to human body movements. From the mechanics of a gently crumpled foil we develop theoretical models and scaling laws for the open circuit voltage and short circuit current and verify via experiments. It is concluded that stretching is the dominant charge generation mechanism with the open circuit voltage and short circuit current scaling as $l^{i}$ and $l^{i-1}(\mbox{d}l/\mbox{d}t)$, respectively with $1\leq i\leq 4/3$ and $l$ the height of the crumple cone.
\end{abstract}

\maketitle
Wearable electronic devices are of interest for personal computing applications such as health diagnostics, identification, communication etc. Energy storage is a challenge for these applications and technologies such as flexible batteries, super capacitors and energy scavengers have been investigated in response to this. Considering that body movements generate $>$ 100 W of power \cite{Source1}, several methods of mechanical energy harvesting have been studied, some more suited for low frequencies than others. Piezoelectric (PZ) based energy harvesters offer promise for flexible wearable electronics with embeds in shoes \cite{Source2,Source3}, fibers \cite{Source4,Source5} and bio implants \cite{Source6,Source7} having been investigated.

\begin{figure}
\centering
\includegraphics[width=3.25 in]{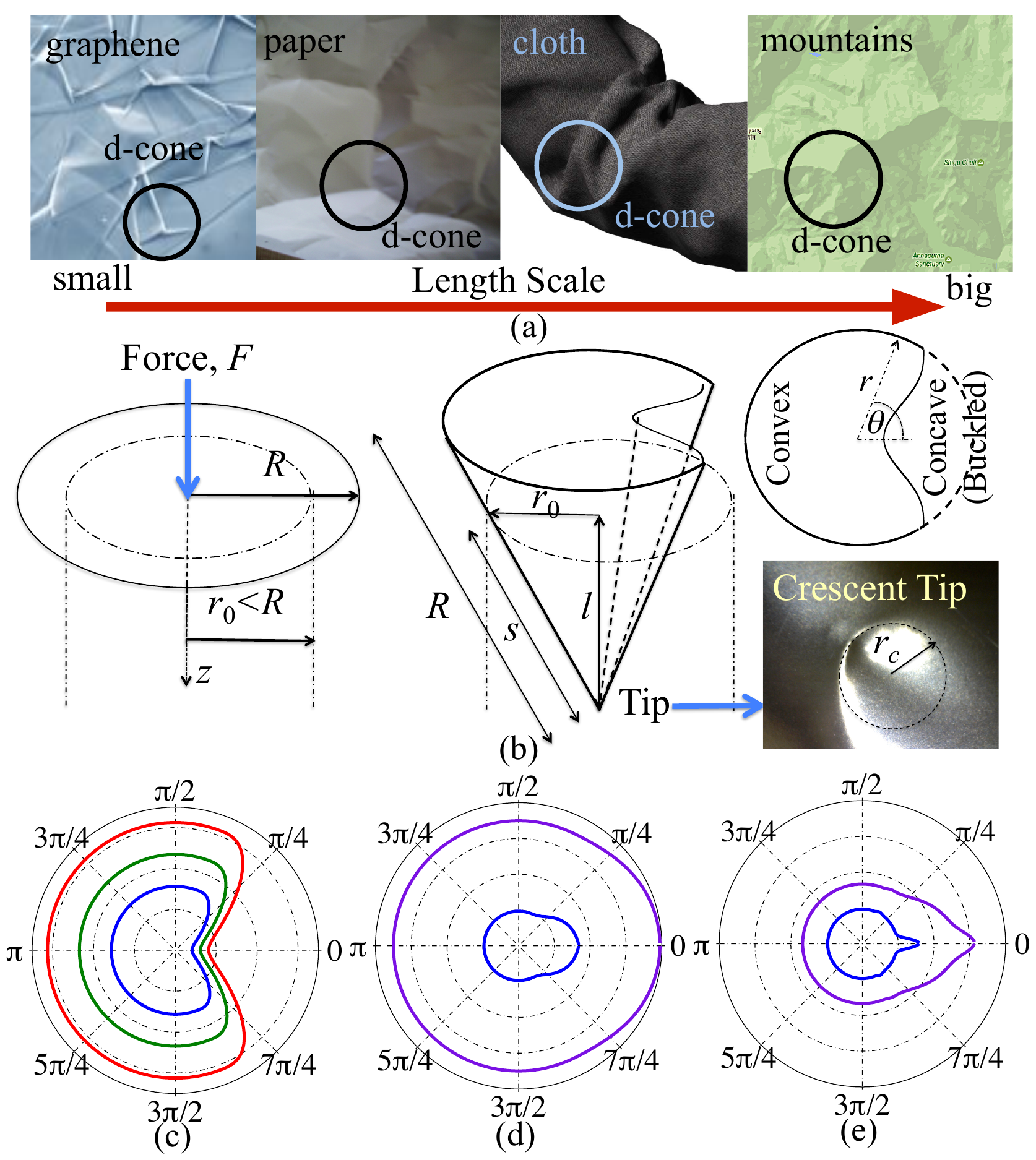}
\caption{(a) Features on a crumpled surface. Picture of graphene \cite{graphene}, mountain range \cite{googlemaps}. (b) Formation of a d-cone. (c) $(r,\theta)$ for the d-cone at $z=l/2$ (blue), $z=3l/4$ (green) and $z=l$ (red). (d) Logarithm of effective stress and (e) charge per unit length developed in the hoop elements at $z=l/2$, with (purple) and without (blue) stretching.}
\end{figure}

In this letter we consider using polymer PZ foils embedded in clothes for sensing and energy harvesting when clothes crumple during activity.  Crumpling can be imagined to be the application of an enclosing boundary of forces to a region on a thin sheet. In response, the sheet will have to achieve mechanical equilibrium without loss of surface area. This results in the formation of the various features seen on a crumple (Fig. 1a). Studies have shown that these features mostly consist of well defined nodes connected by ridges that indicate regions where the sheet has stretched \cite{Cerda1}-\cite{Mora}. The geometry of the nodes appear like swallow-tails \cite{Witten2} and can be emulated by developing a cone (developed cone or d-cone) from a circular sheet without loss of surface area \cite{Cerda1}-\cite{Mahadevan1}, \cite{Witten1}, \cite{Liang}. Ridges connect the vertices of several d-cones to form a network \cite{Witten1}-\cite{Mora}. In this letter we estimate the extractable power from the crumpling of a PZ foil. We develop models for the open circuit voltage and short circuit current during the formation of a d-cone and verify the models by experiments. Studies are performed on polymer PZ foils bonded to a soft pliable cloth to form a composite sheet. A woven cloth offers low mechanical resistance to bending and to forces localized to the length scale of gaps in the weave. Although the mechanics of the sheet depends on the properties of the foil, weave, fabric and the foil-cloth bonding, it is largely determined by the stiffer PZ foil. However, the neutral plane is not the mid plane of the PZ foil, and is offset by a factor dependent on the ratio of the elastic modulus-thickness product of the fabric and foil. Therefore, voltage can be developed across the PZ foil in pure bending. We consider the sheet to have an effective elastic modulus, $E$, and effective thickness $h$ with the Poisson ratio set a constant.

\begin{figure*}
\centering
\includegraphics[width=6.65in]{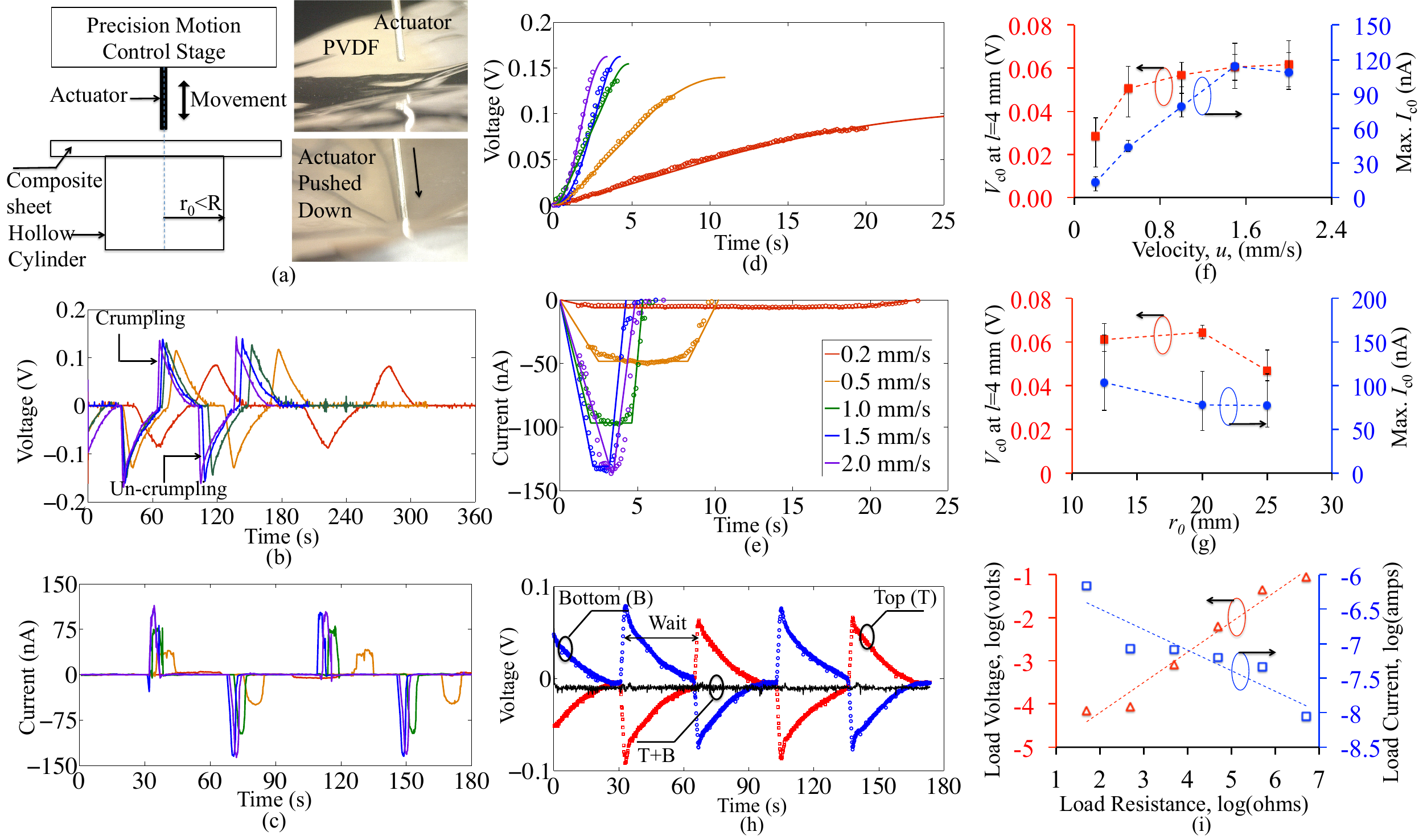}
\caption{(a) Experimental setup (b) Open circuit voltage waveforms and (c) Short circuit current waveforms, for periodic crumpling and uncrumpling at different velocities. (d) Dynamics of open circuit voltage and (e) the dynamics of short circuit current, during crumpling at different velocities. The solid lines indicate the model of Eq. 3 after accounting for the appropriate acceleration of the actuator during the creation of the d-cone. For $u=2$ mm/s, $l\approx 5$mm. The legend is valid for plots (b)-(e). Trend in open circuit voltage and short circuit current measured at maximum displacement and maximum velocity, respectively, with (f) velocity and (g) $r_{0}$. (h) Open circuit voltage waveforms measured from the top and bottom electrodes separately and simultaneously with respect to a common ground. (i) Voltage and current measurements across a load resistor driven by the crumpling of the PVDF foil at $u=2$ mm/s.}
\end{figure*}

A d-cone is experimentally created by forcing a thin circular sheet of radius $R$ into a cylindrical tube of radius $r_{0}<R$ and upto a depth $l$ by applying a point force, $F$, at the center of the sheet as shown in Fig. 1b \cite{Cerda1}-\cite{Mahadevan1}. We consider creating a d-cone with the composite sheet. For gentle bending ($l<h/2$) the mechanics can be described by the Kirchoff theory. The sheet would experience a per unit length shear of $F/2\pi r$ in the region $r\leq r_{0}$ and zero shear elsewhere with both the radial stress and $z$ being zero at $r=r_{0}$. The force required to displace the center of the sheet by $l$ is $F\sim Eh^{3} l/r_{0}^{2}$ where the symbol $\sim$ implies a scaling relation. For the region $r\leq r_{0}$,  the circumfrential and radial stress $\sim F(a_{1}+\mbox{ln}(r_{0}/r))$ and $\sim F\mbox{ln}(r_{0}/r)$, respectively with $a_{1}$ being a constant dependent on the Poisson ratio. For $r>r_{0}$, the circumfrential stress $\sim F(1+(r_{0}/r)^{2})$ and the radial stress $\sim F(1-(r_{0}/r)^{2})$  resulting in a summation that is independent of $r$. As $r$ approaches $r_{0}$, the circumferential stress prevails and the radial stress disappears. However, both are comparable near the center indicating a need for curvature in two directions. Upon continually increasing $l$,  the region around $r=0$ experiences immense stress to beyond the elastic limit thereby permitting the sheet to buckle and fold about this central point. In such a case, two facts determine the geometry of the sheet far from the center. The first is that the energy required to purely bend $\sim Eh^{3}l^{2}/r_{0}^{2}$ while the energy required to stretch $\sim Ehl^{4}/r_{0}^{2}$.  As $h/l$ gets smaller, it is energetically more economical to bend rather than stretch. Therefore, the sheet would localize stretching to the region around the center while experiencing pure bending elsewhere while subject to the constraints of not losing surface area and lying within the hollow cylinder \cite{Cerda1}-\cite{Mahadevan1}, \cite{Witten1}, \cite{Liang}. The second observation is due to Gauss' Theorema Egregium that states that if a surface with zero Gaussian curvature experiences pure bending and no stretching, the final geometry must also retain a zero Gaussian curvature. This implies that the regions of the initially flat sheet that experienced pure bending will have a curvature in one direction only. Both these observations imply that the geometry of a contiuously displaced sheet will be that of a developed-cone (d-cone) with a buckled region as shown in Fig. 1b. Traversing along along $\theta$, the d-cone will touch the rim of the cylinder and fits the envelope of a right angled cone of slant length, $s=(r_{0}^{2}+l^{2})^{1/2}$ until the points where it buckles. In the buckled region, the surface of the d-cone loses contact with the rim and is concave. Fig. 1c shows the plots of $(r,\theta)$ at different $z$ using models developed by Cerda et. al \cite{Cerda1}. In general a hoop of the d-cone can be considered to have curvature $f/r$ with $f$ being a dimensionless function of $\theta$. The circumferential stress along a hoop far from the center has been shown \cite{Mahadevan} to vary as $\sim E(hf/r)^{2}$. Close to the tip of the d-cone, the bending and stretching energies are comparable and the region is morphed to have non-zero Gaussian curvature. Intense stretching can create a plastic deformation leaving a crescent like scar of radius $r_{c}$ as shown in Fig. 1b. The sum of the per unit volume bending and stretching energy in the vicinity of this scar is $\propto E(h/r_{c})^{2}+E (r_{c}/r_{0})^{4}$. The energy sum minimizes for $r_{c}\propto (hr_{0}^{2})^{1/3}$. Clearly the strain and curvature in this region depend on $l$ and a more accurate estimate for $r_{c}$ was shown  to be $r_{c} \sim (r_{0}/l)^{n}(hr_{0}^{2})^{1/3}$ with $n=1/3$ for small deflections and $n=1/2$ for larger deflections \cite{Mahadevan1}. The strain and curvature in this region are $\sim ((l/r_{0})r_{c})^{2}/r_{0}^{2}$ and $\sim (l/r_{0})(1/r_{c})$, respectively. For $l<r_{0}$ we consider stretching to be the dominant contributor to stress in the tip. 

To determine the charge in the d-cone due to crumpling, the tangential stress, $\sigma_{tc}$, and shear stress, $\tau_{zc}$, in the finite elements need to be defined. The radial and circumferential stresses due to bending and stretching contribute to $\sigma_{tc}$. If the PZ coefficients are defined as $(d_{31}=d_{32}, d_{15}=d_{24})$ with the 3-direction being normal to the surface of the foil and the 1-direction and 2-direction being tangential to the foil surface and perpendicular to each other, the charge per unit area developed on the PZ foil is $d_{31}\sigma_{tc}+d_{15}\tau_{zc}$. Ignoring constants in the prefactor, $\sigma_{tc}$ and $\tau_{zc}$ in the finite elements of the PZ foil scale as,
\begin{align}
\sigma_{tc}&\sim
\begin{cases}
Ehl/r_{0}^{2}(a_{1}+\mbox{ln}(r_{0}/r))&\mbox{if}\hspace{0.035 in}l\lesssim h, r\leq r_{0}\\
Ehl/r_{0}^{2} &\mbox{if}\hspace{0.035 in}l\lesssim h, r>r_{0}\\
E(h^{2/3}l^{i}/r_{0}^{2/3+i})&\mbox{if}\hspace{0.035 in}l>h, r\lesssim (hr_{0}^{2})^{1/3}\\
E(hf/r)^{2}&\mbox{if}\hspace{0.035 in}l>h, r>(hr_{0}^{2})^{1/3}
\end{cases}\nonumber \\
\tau_{zc}&\sim
\begin{cases} 
E(hl/r_{0}^{2})(h/r) &\mbox{if}\hspace{0.035 in}r\leq r_{0}\\
0&\mbox{if}\hspace{0.035 in}r>r_{0}
\end{cases}
\end{align}
Fig. 1d compares the logarithm of the effective stress in a hoop of the d-cone with (purple) and without (blue) the presence of stretching. Here $i=2(1-n)$ with $i=4/3$ when $l/r_{0}\lesssim 0.1$,  $i=1$ when $l/r_{0}>0.1$.  There is no smooth transition in the model between these two cases.

The charge developed in each elemental section of the d-cone is $(d_{31}\sigma_{tc}+d_{15}\tau_{zc})(r\mbox{d}r\mbox{d}\theta)/(\mbox{sin}(\mbox{tan}^{-1}(\mbox{d}r/\mbox{d}z)))$. Due to the conservation of charge, the charge developed in each finite element of the d-cone can be summed to obtain the total charge. Several approximations are made. First, for the case of $l\lesssim h$ where the bending is extremely gentle, it is assumed that $\mbox{tan}^{-1}(\mbox{d}r/\mbox{d}z) \sim \pi/2$ for all $r$. Second, since the geometry near the tip of the d-cone is not easily defined, this region is considered to have a surface area $\sim (h^{1/3}r^{2/3})^{2}$. Finally, since the concave region experiences the opposite kind of bending stress as compared to the convex region, the charge developed in these regions would be of the opposite sign with the sign depending on the PZ polarization direction. Therefore the charge on the concave region would attempt to nullify the charge in the convex region via surface currents with only the excess remaining charge contributing to the voltage. As the varying curvature makes the exact calculation of charge difficult, we model the excess charge developed in the region of $r>>h^{1/3}r^{2/3}$ as a constant factor times the charge developed in the convex region. All these assumptions are justified by experimental results. Using these arguments, the charge developed in the d-cone is seen to vary as,
\begin{equation}
q_{c}\sim
\begin{cases}
Ed_{31}lh+b_{1}Ed_{15}lh(h/r_{0})((R/r_{0})-1)&\mbox{if}\hspace{0.035 in} l\leq h\\
Ed_{31}l^{i}h^{4/3}/r_{0}^{i-2/3}+b_{2}Ed_{15}h^{2}(ls/r_{0}^{2})\\
\hspace{0.1 in}+b_{3}Ed_{31}h^{2}(s/r_{0})\mbox{ln}((R/s)(r_{0}/h)^{1/3}))
\hspace{0.1 in} &\mbox{if}\hspace{0.035 in} l>h
\end{cases}
\end{equation}
Here $b_{1}, b_{2}, b_{3}$ are constant coefficients. If the neutral plane of the sheet lies very close to the mid plane of the PZ foil, $b_{3}$ can be expected to be very small. Fig. 1e compares the logarithm of the per unit length effective charge developed in the hoop elements of the d-cone with (purple) and without (blue) stretching. The total energy content of the d-cone is $g_{c}=q_{c}^{2}/2C_{c}$ with $C_{c}\sim \epsilon R^{2}/h$ being the capacitance and $\epsilon$ the permittivity. With periodic crumpling and uncrumpling, $l$ becomes a function of time $t$, and the estimate of harvestable power is $P_{c}=\mbox{d}g_{c}/\mbox{d}t$. This estimate can be decomposed into the short circuit current, $I_{c0}=\mbox{d}q_{c}/\mbox{d}t$ and open circuit voltage, $V_{c0}=q_{c}/C_{c}$. In our case the thin foil ($h/r_{0}\approx 10^{-3}$) is moderately crumpled into a d-cone ($l/r_{0}<0.33$). Defining $s \sim r_{0}$, $i=1$ in Eq. 2 and noting that bending near the tip is less than streching , $V_{c0}$ and $I_{c0}$ scale as,
\begin{align}
V_{c0}&\sim E(h^{2}/C_{c})(d_{31}l/(h^{2}r_{0})^{1/3}+k_{1}d_{31}\mbox{ln}(R/(hr_{0}^{2})^{1/3}) \nonumber \\
&\hspace{0.1 in}+k_{2}d_{15}(l/r_{0}))  \nonumber \\
I_{c0}&\sim E(h^{2}/r_{0})(d_{31}(r_{0}/h)^{2/3}+k_{3}d_{15})(\mbox{d}l/\mbox{d}t)
\end{align}
Here $(k_{1},k_{2}, k_{3})$ are constant coefficients.

The experimental setup to evaluate these models for the d-cone is shown in Fig. 2. Experiments were performed using a 52 $\mu$m thick poly vinyldenefluoride (PVDF) \cite{PVDF} based PZ foil with with electrodes on both sides and bonded to a soft cloth-plaster of thickness 250 $\mu$m to form the composite sheet. The PVDF had $E=5$ GPa, $d_{31}=$5 pCm$^{-2}$/Nm$^{-2}$ and $\epsilon=$88.5$\times$10$^{-12}$ F/m. An actuator was used to push the center of this sheet ($\sim 50$ mm) into a cylindrical tube of radius $r_{0}=(12.5, 20, 25)$ mm at constant velocity $u$ and upto $\mbox{max}(l)\approx 4$ mm to create the d-cone (crumpling). To achieve this actuator accelerated appropriately. After a 30 s wait, the actuator was pulled up with velocity $u$ and to the initial position thereby allowing the sheet to relax (uncrumpling). The open circuit voltage, $V_{c0}$ and short circuit current, $I_{c0}$ were measured during periodic crumpling and uncrumpling events separated by the wait period.  Fig. 3b and Fig. 3c show the dynamics of $V_{c0}$ and $I_{c0}$ for different $u$. The voltage rises or falls sharply during crumpling or uncrumpling resulting in current pulses. During the wait period, charge leaks away through some parasitic load resistance and the voltage gradually moves to zero. This leakage is not visibly observed in the $I_{c0}$ waveforms but since $C_{c}\approx 5$ nF, and leakage time constant is $\approx$ 20s, this load is $\approx$ 4 G$\Omega$ and of the same order as the PZ impedance considering the event dynamics is $\approx 1 s$. As the equivalent circuit is defined by the PZ capacitance and the load in series, the measured voltage is not the true open circuit voltage but a frequency dependent factor times $V_{0c}$. Therefore although $V_{oc}$ does not depend on $u$, the voltage across the load would do so. Fig. 3d and Fig. 3e show the transients of $V_{c0}$ and $I_{c0}$ during crumpling. The solid line shows the model based on Eq. 3 $(k_{1}=k_{2}=k_{3}=1)$ and accounts for the acceleration of the actuator. The almost linear dependence of $V_{c0}$ on $l$ in accordance with Eq. 3 is observed. A point of note is that the factor $i$ in Eq. 3 must always be $\geq 1$. If $i<1$,  $I_{c0}$ would contain a $l^{i-1}$ term resulting in transients that would not conform with experiment. Therefore the models as developed are in good corroboration with experiment. Fig. 3f shows the trend in maximum $V_{c0}$ and $I_{c0}$ with $u$. According to Eq. 3, $V_{c0}$ is independent of $u$. However as discussed above, the presence of the load impedance makes the voltage measured depend on $u$. At higher $u$, the measured voltage $\approx V_{0c}$. According to Eq. 3, $I_{c0}$ is expected to linearly increase with $u$ as observed. Fig. 3g shows a gradual decreasing trend of $V_{c0}$ and $I_{c0}$ with increasing $r_{0}$ tending towards agreement with the $r_{0}^{-1/3}$ dependence predicted in Eq. 3. Fig. 3h shows  experiments on a stand alone PZ foil. The voltage waveform from the top and bottom electrodes were measured separately and simultaneously with a common ground. The open circuit voltage is the difference of these two waveforms. In this case, the neutral plane lies in the middle of the PZ foil, and pure bending (excepting shear) should produce no open circuit voltage. However, since the crumpling and uncrumpling events produce oppositely going voltage pulses, it is indicative of stretching being the major contributor. The sum of these waveforms indicates the contribution of bending (black). Fig. 3i shows the load current and load voltage with the PZ foil driving varying loads.

In summary, the stretching near the tip contributes more to charge generation as compared to bending elsewhere. Fast, strong crumpling creating multiple d-cones as opposed a single large d-cone offers best use of area. The active area scales as $(h/r_{0})^{2/3}$ while the voltage and current scale as $l^{i}$ and $l^{i-1}(\mbox{d}l/\mbox{d}t)$, respectively with $1\leq i\leq4/3$. Efficiency can be improved by disconnecting the concave side from the convex if the locations are predictable \cite{sanjiv}. For strong crumpling, the contribution of the ridge must be considered \cite{Witten1}-\cite{Mora}. The ridge stretches by the same order as the tip of the d-cone. Since the distance between the tips of two d-cones is $2r_{0}$, the charge developed over the area $(2r_{0}) (h^{1/3}r_{0}^{2/3})$ is $\sim Ed_{31}hl$. Although difficult, a ridge could be emulated by the use of two actuators (the distance between being $2r_{0}$) creating two d-cones simultaneously. For random crumpling, the probability density function for $r_{c}$ and ridge length can be used to predict $\left<r_{0}\right>$ and mean harvestable power \cite{dconestatistics}, \cite{Sultan}.

%

\end{document}